\documentclass[12pt]{article}
\usepackage{amsmath,amssymb}

\def\Dslash{\not{\hbox{\kern-4pt $D$}}}
\def\dslash{\kern-4pt \not{\hbox{\kern-2pt $\partial$}}}
\def\pslash{\not{\hbox{\kern-2pt p}}}

\begin{document}

\providecommand{\V}{\mathcal{V}} \providecommand{\rh}{\hat{r}}

\begin{titlepage}

\begin{flushright}
                             \texttt{hep-th/0405276}\\
                             DSF--11--2004
\end{flushright}
\vspace{1.5cm}

\begin{center}
\renewcommand{\baselinestretch}{1.8}\normalsize
\textbf{\LARGE Supersymmetric Dissipative Quantum Mechanics from Superstrings}
\\[1.5cm]

{\large Luigi~Cappiello and Giancarlo~D'Ambrosio}\\[0.5cm]

\renewcommand{\baselinestretch}{1.2}\normalsize
\emph{Dipartimento di Scienze Fisiche, Universit\`{a} di Napoli
``Federico II''}\\
\emph{and}\\
\emph{INFN -- Sezione di Napoli,}\\
\emph{Via Cintia, 80126 Napoli, Italy}\\[12pt]
\emph{E-mails:} \texttt{Luigi.Cappiello,Giancarlo.Dambrosio@na.infn.it}\\[1.5cm]
\end{center}

\begin{abstract}
Following the approach of Callan and Thorlacius applied to the
superstring, we derive a supersymmetric extension of the non-local
dissipative action of Caldeira and Leggett. The dissipative term
turns out to be invariant under a group of superconformal
transformations. When added to the usual kinetic term, it provides
an example of supersymmetric dissipative quantum mechanics. As a
by-product of our analysis, an intriguing connection to the
homeotic/hybrid fermion model, proposed for CPT violation in
neutrinos, appears.
\end{abstract}
\end{titlepage}

\section{Introduction}

Dissipative classical and quantum systems have constantly received
attention, along the years, by both theoreticians and
experimentalists. On  one side, this subject is strictly related
to other interesting questions about decoherence, and
non-equilibrium physics of open systems with a small number of
degrees of freedom interacting with an external environment. On
the other side, it could be relevant in explaining experimental
results in the physics of mesoscopic systems.

A model describing the dissipative behavior of a quantum
mechanical system was given long ago by Caldeira and Leggett (CL)
\cite{Caldeira:1982uj}. Briefly speaking, the CL model of
dissipative quantum mechanics consists in considering a quantum
system in interaction with a bath formed by an infinite number of
harmonic oscillators. If the interaction is linear and if the
spectral function of the frequencies of the oscillators satisfies
what is called the Ohmic condition, the bath produces a non-local
effective term which introduces dissipation in the dynamics of the
quantum system. Some time later, Callan and Thorlacius
\cite{Callan:1989mm} found an interesting derivation of the CL
model in string theory. They showed that open bosonic string
theory provides a way of generating the CL non-local term as the
effective action describing the interaction of the string
end-point with the oscillation modes of the string. The original
derivation of \cite{Callan:1989mm} follows methods developed in
\cite{Fradkin:1985qd} and \cite{Abouelsaood:gd}, and uses the
formalism of boundary states \cite
{Polchinski:1987tu},\cite{Callan:1988wz}. A somewhat simpler
derivation of their result, making use of a (constrained)
functional integration was given in \cite{Iengo:1997hz}.

In this paper we show that similar reasonings, applied to the
fermionic string, produce a new non-local effective action which
is the fermionic analogous of the CL non-local term. When
contributions from the bosonic and fermionic coordinates of the
string are considered, one gets both the CL term and its fermionic
companion. We find that they are separately invariant under
$SL(2,R)$ subgroup of the full reparametrization group of the
string, and that when they are put together the resulting
expression is supersymmetric. It can be explicitly written in
superspace formalism and turns out to be invariant under
superconformal transformations. When it is added to a
supersymmetric kinetic term, it provides the first example of a
supersymmetric dissipative quantum mechanics, which is the main
result of this paper.

The plan of the paper is as follows. In Section 2 we give a short
review of the Caldeira-Leggett construction of the non-local
action which describes dissipation in a quantum mechanical system.
In Section 3 derive the analogous non-local action for a fermionic
system, following the method of constrained functional
integration, of ref.\cite{Iengo:1997hz}. In Section 4, another
derivation, which is closer in spirit to the original derivation
of \cite{Callan:1989mm} is given using the boundary state
formalism. In Section 5, we show that the non-local fermionic
dissipative term is invariant under the $SL(2,R)$ subgroup of the
full reparametrization group of the string. Furthermore, we show
that the bosonic (CL) and fermionic terms are parts of a
superconformal invariant action which we write it explicitly in
superspace. In Section 6, the dissipative non-local term is added
to a supersymmetric kinetic term to form the action of a
supersymmetric dissipative quantum mechanical system. Some further
discussions and speculations, including a possible interpretation
of the homeotic/hybrid fermionic field theory model proposed and
criticized respectively  in \cite{Barenboim:2002tz} and
\cite{Greenberg:2003ks},  can be found in the conclusive Section.

\section{The Caldeira-Leggett model of dissipative quantum mechanical systems}

The simplest example of classical dissipative system is the particle moving
 in the presence of friction. Its equation of motion is
\[
M\frac{d^{2}q}{dt^{2}}+\eta \frac{dq}{dt}+\frac{dV_{0}(q)}{dq}=0
\]
where $V_{0}(q)$ is the potential and $\eta $ gives the strength
of the friction force. As is well known there is no action whose
variation can produce an equation of motion containing the term
proportional to $\eta $, and thus there is no hamiltonian
structure and no canonical quantization procedure for this system.

According to Caldeira and Leggett \cite{Caldeira:1982uj}, quantum
dissipative mechanical behavior can be obtained from a microscopic model
consisting in coupling the mesoscopic system with an environment modelled as
an infinite set of harmonic oscillators. The CL microscopic action of the
particle on the line can then be written as follows:
\begin{eqnarray}
S[q,\{x_{\alpha }\}] &=&\int dt[\frac{1}{2}M\left(
\frac{dq}{dt}\right)
^{2}-V_{0}(q)+\sum_{\alpha }\left( \frac{1}{2}m_{\alpha }\left( \frac{%
dx_{\alpha }}{dt}\right) ^{2}-\frac{1}{2}m_{\alpha }\omega
_{\alpha
}^{2}x_{\alpha }^{2}\right)  \label{microcl} \\
&&-q\left( \sum_{\alpha }C_{\alpha }x_{\alpha }+F_{ext}(t)\right)
-\sum_{\alpha }\kappa \frac{C_{\alpha }^{2}q^{2}%
}{2m_{\alpha }\omega _{\alpha }^{2}}],  \nonumber
\end{eqnarray}
where $q(t)$ is the dynamical variable of the mesoscopic system and $%
\{x_{\alpha }(t)\}$ is the infinite set of harmonic oscillators
representing the environment. Without lack of generality one can
consider
that the system is coupled linearly to each oscillator with strength $%
C_{\alpha }$. $F_{ext}(t)$ represents an external force. The last term in (%
\ref{microcl}) has to be introduced to take into account a
renormalization effect to the classical potential $V_{0}(q)$. The
classical equations of motion of (\ref{microcl}) are
\begin{eqnarray}
M\frac{d^{2}q}{dt^{2}} &=&-\frac{dV_{0}(q)}{dq}-\sum_{\alpha }\left(
C_{\alpha }x_{\alpha }+\kappa \frac{C_{\alpha }^{2}q}{m_{\alpha }\omega
_{\alpha }^{2}}\right) +F_{ext}(t)  \label{eqmicrocl} \\
m_{\alpha }\frac{d^{2}x_{\alpha }}{dt^{2}} &=&-m_{\alpha }\omega _{\alpha
}^{2}x_{\alpha }-C_{\alpha }q  \nonumber
\end{eqnarray}
Fourier transforming, solving for $x_{\alpha }$ in the second
equation and plugging into the first equation, one obtains
\begin{equation}
-M\omega ^{2}\widetilde{q}(\omega )=-\widetilde{\frac{dV_{0}}{dq}}(\omega )+%
\widetilde{F}_{ext}(\omega )+\widetilde{K}(\omega )\widetilde{q}(\omega )
\label{fourier}
\end{equation}
where
\[
\widetilde{K}(\omega )=-\sum_{\alpha }\frac{C_{\alpha }^{2}}{m_{\alpha
}\omega _{\alpha }^{2}}\dfrac{\omega ^{2}}{\omega ^{2}-\omega _{\alpha }^{2}}
\]
One gets an imaginary part $\widetilde{J}(\omega )=\mathrm{Im}$\textrm{\ }$%
\widetilde{K}(\omega )$ by the usual rule $\omega \rightarrow \omega
+i\varepsilon $, which shifts the poles from the real axis:
\[
\widetilde{J}(\omega )=\frac{\pi }{2}\sum_{\alpha }\frac{C_{\alpha }^{2}}{%
m_{\alpha }\omega _{\alpha }^{2}}\delta (\omega -\omega _{\alpha }).
\]
(Notice that expression of $\widetilde{J}(\omega )=\mathrm{Im}$ $\widetilde{K%
}(\omega )$, is independent of the presence of the last term in (\ref
{microcl}), which only affects the real part of $\widetilde{K}(\omega )$. On
the other hand, $\widetilde{K}(\omega )$ is related to $\widetilde{J}(\omega
)$ by a subtracted dispersion relation when the renormalization correction
term is taken into account. One can then show that in that case $\widetilde{J%
}(\omega )=\mathrm{Im}$ $\widetilde{K}(\omega )$ is the dominant
contribution with respect to $\mathrm{Re}$ $\widetilde{K}(\omega )$ for $%
\omega $ below some characteristic frequency of the system.) If the
parameters of the bath of harmonic oscillators satisfy the Ohmic condition
\begin{equation}
\widetilde{J}(\omega )=\eta \omega ,  \label{ohm}
\end{equation}
then the eq.(\ref{fourier}) reduces to
\begin{equation}
-M\omega ^{2}\widetilde{q}(\omega )=-\widetilde{\frac{dV_{0}}{dq}}(\omega )+%
\widetilde{F}_{ext}(\omega )+i\eta \omega \widetilde{q}(\omega )
\label{fourierdiss}
\end{equation}
\textit{i.e.} an effective friction term is generated in the
equation of motion of the particle.

Quantum-mechanically, the interaction of the particle with the oscillator
bath produces a non-local effective action term, once the Euclidean
functional integration on the $x_{\alpha }(t)$ is performed. The result is
\[
S_{CL}[q]=-\int dt_{1}\int dt_{2}q(t_{1})\alpha (t_{1}-t_{2})q(t_{2}),
\]
where the non local coupling coefficient is given by
\[
\alpha (t)=\int_{0}^{\infty }\frac{d\omega }{2\pi }\widetilde{J}(\omega
)e^{-\omega |t|}.
\]
In case of Ohmic dissipation (\ref{ohm}), one obtains the
Caldeira-Leggett non local term \cite{Caldeira:1982uj}
\begin{equation}
S_{CL}[q]=\frac{\eta }{2\pi }\int dt_{1}\int dt_{2}\frac{q(t_{1})q(t_{2})}{%
(t_{1}-t_{2})^{2}}=\frac{\eta }{4\pi }\int dt_{1}\int dt_{2}\frac{%
(q(t_{1})-q(t_{2}))^{2}}{(t_{1}-t_{2})^{2}},
\end{equation}
The introduction of the CL term has been seminal in the study of
dissipative quantum mechanics. A general reference on the subject
is \cite{weiss}.

Working in two-dimensional bosonic string theory, Callan and
Thorlacius \cite {Callan:1989mm} found a connection between the
so-called boundary states and one-dimensional dissipative quantum
mechanics. Roughly speaking, boundary states are a sort of
coherent states in the Fock space of string theory, which
implement operatorially open string boundary conditions. (On this
point we shall be much more explicit in Section 4.) The main idea
is that, while evolving in time, the open string end-point behaves
like a point particle interacting with the infinite set of
harmonic oscillator formed by the string modes. As such, its
dynamics is dissipative. In the next Section we illustrate how a
dissipative non-local action for the dynamics of the string
end-point, can be obtained by functional-integrating out the open
string oscillator modes.

\section{A dissipative term from the fermionic string}

Our first derivation of the fermionic non-local dissipative \ term
follows the approach used in ref. \cite{Iengo:1997hz} to derive
the CL lagrangian from the open bosonic string. It consists in the
direct evaluation of a constrained open string partition function,
where the string fields are bound to assume some fixed boundary
values at one of the string end-points.

Let us begin by writing the action of the fermionic string as \cite{Green:sp}
\begin{equation}
\mathcal{S}=-\frac{1}{4\pi \alpha ^{\prime }}\int_{-T}^{T}d\tau
\int_{0}^{\pi }d\sigma \left( \partial _{a}X_{\mu }\partial ^{a}X^{\mu }-i%
\overline{\psi }_{\mu }\rho _{a}\partial ^{a}\psi ^{\mu }\right) .
\end{equation}
Here the $X=X^{\mu }(\tau ,\sigma )$ are the bosonic coordinates
of the string, the $\psi =\psi ^{\mu }(\tau ,\sigma )$, are
massless Majorana spinors on the string world-sheet, and the $\rho
_{a},$ $a=0,1$ are two-dimensional Dirac matrices. In the
following, we shall momentarily neglect the bosonic part of the
action, and consider only a single fermionic field. Using
light-cone coordinates, the fermionic action can be written
\[
\mathcal{S}=\frac{i}{4\pi \alpha ^{\prime }}\int d\tau d\sigma \left( \psi
_{-}\partial _{+}\psi _{-}+\psi _{+}\partial _{-}\psi _{+}\right) .
\]
Boundary conditions can be written in terms of $\psi _{-}(\tau
,\sigma )$ defined on the doubled interval $\in \left( 0,2\pi
\right) $, with $\psi _{+}(\tau ,\sigma )=\psi _{-}(\tau ,2\pi
-\sigma ).$ The Ramond and Neveu-Schwartz sectors correspond to
$\psi _{-}(\tau ,\sigma )$ periodic or anti-periodic on the
doubled interval. The corresponding mode expansions are of the
form
\begin{equation}
\psi _{\pm }(\tau ,\sigma )=\sum_{p=-\infty }^{\infty }\psi _{p}(\tau
)e^{\pm ip\sigma },\text{ }
\end{equation}
where $p$ is integer in the Ramond sector and half-integer in the
Neveu-Schwartz case. We then perform a Wick rotation in time $\tau
\rightarrow it$ and Fourier expand also in $t\in (-T,T)$, obtaining
\[
\psi _{\pm }(t,\sigma )=\sum_{p=-\infty }^{\infty }\sum_{k=-\infty }^{\infty
}\psi _{p,k}e^{i\left( \frac{2\pi k}{T}t\pm p\sigma \right) }
\]
with $(\psi _{p,k})^{\ast }=\psi _{-p,-k}$, following from the reality of $%
\psi _{\pm }(t,\sigma ).$
Substituting these expansions into the
Euclidean action, we obtain after some straightforward algebra
\[
S_{E}=\frac{T}{\alpha ^{\prime }}\sum_{p=-\infty }^{\infty
}\sum_{k=0}^{\infty }(p+\frac{2\pi i}{T}k)\psi _{-p,-k}\psi _{p,k}.
\]
Consider now the mode expansion of the assigned boundary value of
the fermionic field, \textit{i.e},
\[
\psi _{\pm }(t,0)=\sum_{p=-\infty }^{\infty }\sum_{k=-\infty }^{\infty }\psi
_{p,k}e^{\pm i\left( \frac{2\pi ik}{T}t\right) }=\sum_{k=-\infty }^{\infty
}\chi _{k}e^{\pm i\left( \frac{2\pi ik}{T}t\right) }\equiv \chi (t)
\]
The one-dimensional effective action for the string end-point
fermionic variable $\chi (t)$, is obtained by computing the
following constrained functional integral, which is the annulus
diagram of the open string, with the assigned boundary conditions
\begin{equation}
e^{-S_{eff}(\chi )}=\mathcal{N}\int \left[ \prod_{k>0}\prod_{p}d\psi
_{-p,-k}d\psi _{p,k}\right] \prod_{k}d\psi _{0,k}\,e^{-\mathcal{S}%
_{E}}\,\prod_{k}\delta \left( \sum_{p}\psi _{pk}-\chi _{k}\right) ,~
\nonumber
\end{equation}
The constraints, given by delta functions, are easily solved,
giving $\psi _{0,k}=\chi _{k}-\sum_{p\neq 0}\psi _{p,k}$ . The
resulting
integrals for each $k$ are then of the form (we momentarily drop the factor%
\textit{\ }$T/\alpha ^{\prime }$ from\textit{\ }$\mathcal{S}_{E}$)
\begin{eqnarray}
e^{-S(\chi _{k})} &=&\mathcal{N}e^{\frac{2\pi ik}{T}\chi _{-k}\chi _{k}}\int
\prod_{p}d\psi _{-p,-k}d\psi _{p,k}  \nonumber \\
&&\exp \left\{ \frac{2\pi ik}{T}\left[ \chi _{-k}\left( \sum_{p\neq 0}\psi
_{p,k}\right) +\left( \sum_{p\neq 0}\psi _{-p,-k}\right) \chi _{k}\right]
\right. +  \label{expact} \\
&&\left. \sum_{p^{\prime }\neq 0}\sum_{p\neq 0}\psi _{-p^{\prime },-k}\left[
\frac{2\pi ik}{T}+\delta _{pp^{\prime }}(p+\frac{2\pi ik}{T})\right] \psi
_{p,k}\right\} ,  \nonumber
\end{eqnarray}
All integrals in (\ref{expact}) are Gaussian and can be done by
completing the squares. Eventually, for each mode $\chi _{k}$ one
obtains \noindent
\begin{equation}
S(\chi _{k})=\frac{1}{\sum_{p=0}1/a(p)},  \label{Sk}
\end{equation}
with
\begin{equation}
a(p)=\left( p+\frac{2\pi ik}{T}\right) .  \label{apf}
\end{equation}
The sum (\ref{Sk}) is not converging (but in a distributional
sense) for $a(p)$ given in (\ref{apf}), nevertheless, one could
regularize it by adding, for instance, a term $\varepsilon p^{2}$
to $a(p)$, which would make it converging, and then passing to the
$\varepsilon \rightarrow 0$ limit after having summed (using
residues). One obtains
\begin{equation}
S(\chi _{k})=\frac{i}{\pi }\tanh (\frac{2\pi ^{2}k}{T})\chi _{-k}\chi _{k}
\label{sdiscrf}
\end{equation}
Alternatively, referring to \cite{Iengo:1997hz} for the details,
one can introduce some dimensional parameter into the action to
define a continuum limit, which corresponds to the case where the
inner circular border of the annulus shrinks to zero, i.e. to the
disk diagram of the string, and replace the sum over $p$ by an
integral, to get
\begin{equation}
S(\chi _{k})=\frac{1}{\int_{0}^{\infty }dx/\left( \frac{2\pi ik}{T}+x\right)
}\chi _{-k}\chi _{k}=\frac{i}{\pi }\,sign(k)\,\chi _{-k}\chi _{k}.
\label{scontf}
\end{equation}
\noindent where the integral has to be considered as principal
value. It easily seen that the two expressions (\ref{sdiscrf}) and
(\ref{scontf}) tend to coincide for high values of $k$, i.e. at
high frequencies, where the continuum limit is obviously a good
approximation of the discrete sum, while they differ at low
frequencies.

We get the effective action on the circle of length $2T$, corresponding to
finite temperature, by summing on $k$, the various contributions in (\ref
{scontf}). However, when dealing with fermions we can consider periodicity
or anti-periodicity in Euclidean time too, corresponding to integer or
half-integer values of $k$. Using string world-sheet dualiy, which exchanges
the open and closed string channel, (anti-)periodicity in time of the open
string fermions will correspond to closed string fermions (anti-)periodic in
the string coordinate $\sigma $.

The resulting action for the theory on the circle, restoring the factor $%
T/\alpha ^{\prime }$, takes the two different non-local expressions:
\begin{equation}
S(\chi )=\dfrac{1}{4\pi \alpha ^{\prime }}\frac{1}{T}\text{P}%
\int_{-T}^{T}dt\int_{-T}^{T}dt^{\prime }\chi (t)\chi (t^{\prime })\cot [\pi
(t-t^{\prime })/T],  \label{fulldissp}
\end{equation}
for fermions periodic in Euclidean time, and
\begin{equation}
S(\chi )=\dfrac{1}{4\pi \alpha ^{\prime }}\frac{1}{T}\text{P}%
\int_{-T}^{T}dt\int_{-T}^{T}dt^{\prime }\frac{\chi (t)\chi (t^{\prime })}{%
\sin [\pi (t-t^{\prime })/T]},  \label{fulldissa}
\end{equation}
for antiperiodic ones. The symbol P indicates that the integral is
principal value.

Both expressions (\ref{fulldissp}) and (\ref{fulldissa}) lead to the same
non local action in the zero temperature $T\rightarrow \infty $, \emph{i.e.}
\begin{equation}
S(\chi )=\frac{1}{4\pi ^{2}\alpha ^{\prime }}\text{P}\int_{-\infty }^{\infty
}dt\int_{-\infty }^{\infty }dt^{\prime }\frac{\chi (t)\chi (t^{\prime })}{%
(t-t^{\prime })}.  \label{fermidissi}
\end{equation}
This term is the analogue, for the anticommuting variable $\chi
(t)$ of the CL dissipative term for the commuting variable $q(t)$.

For the sake of comparison, we summarize here how the standard CL
action was obtained in ref.\cite{Iengo:1997hz} starting from the
bosonic term of the string action. The expression of $a(p)$ is
different from (\ref{apf}), because the bosonic term is quadratic
in the derivatives and one gets
\[
a(p)=\left( p^{2}+\left( \frac{2\pi k}{T}\right) ^{2}\right) .
\]
The corresponding sum in (\ref{Sk}) is now converging and using
residues one obtains
\begin{equation}
S(q_{k})=\frac{2\left( \frac{2\pi k}{T}\right) ^{2}}{1+\pi \left( \frac{2\pi
k}{T}\right) \coth (\frac{2\pi ^{2}k}{T})}q_{-k}q_{k}
\end{equation}
where the $q_{k}$ are the Fourier modes of the boundary conditions
of the bosonic field at $\sigma =0$. In the continuum limit the
sum can be replaced by an integral and one gets
\cite{Iengo:1997hz}
\begin{equation}
S(q_{k})=-\frac{T}{4\pi \alpha ^{\prime }}\frac{1}{\int_{0}^{\infty
}dx/\left( \left( \frac{2\pi k}{T}\right) ^{2}+x^{2}\right) }q_{-k}q_{k}=-%
\frac{1}{\alpha ^{\prime }}\,|k|\,q_{-k}q_{k}.  \label{Scontf}
\end{equation}
Summing over all the integer values $k$, one obtains the CL non local action
on the circle
\begin{equation}
S(q)=-\frac{1}{16\alpha ^{\prime }}\frac{1}{T^{2}}\text{P}%
\int_{-T}^{T}dt\int_{-T}^{T}dt^{\prime }\frac{q(t)q(t^{\prime })}{(\sin [\pi
(t-t^{\prime })/T])^{2}},  \label{fulldissbos}
\end{equation}
which in the limit $T\rightarrow \infty $, reduces to the CL
dissipative term
\begin{equation}
S(q)=-\frac{1}{16\pi ^{2}\alpha ^{\prime }}\text{P}\int_{-\infty }^{\infty
}dt\int_{-\infty }^{\infty }dt^{\prime }\frac{q(t)q(t^{\prime })}{%
(t-t^{\prime })^{2}},  \label{bosodissi}
\end{equation}
with dissipation constant $\eta $ proportional to the inverse of
$\alpha ^{\prime }$.

\section{\protect\bigskip Derivation using boundary states}
The original derivation of the Caldeira-Leggett non-local action
in string theory was done in \cite{Callan:1989mm} using the
formalism of boundary states. It is then not surprising that the
fermionic non local action can be analogously obtained using
fermionic string boundary states. Let us first summarize the
method in the bosonic case \cite{Abouelsaood:gd}. (In this Section
we take $\alpha ^{\prime }=2$, restoring it only in the final
formulas.) A single closed bosonic string coordinate can be
expanded in modes as follows
\[
X(\tau ,\sigma )=q-2ip\tau +\sum_{m\neq 0}\frac{1}{\sqrt{|m|}}\left(
a_{m}e^{-m(\sigma +i\tau )}+\widetilde{a}_{m}e^{-m(\sigma -i\tau )}\right) ,
\]
where the only non vanishing commutation relations of the left and
right
moving creation and annihilation are $\left[ a_{m},a_{n}^{\dagger }\right] =%
\left[ \widetilde{a}_{m},\widetilde{a}_{n}^{\dagger }\right] =\delta _{mn}$,
and $a_{n}^{\dagger }=a_{-n}$, $\widetilde{a}_{n}^{\dagger }=\widetilde{a}%
_{-n}$ The boundary state $|x,\overline{x}>$ is a coherent state satisfying
the equations
\[
\begin{array}{lll}
(a_{m}^{\dagger }+\widetilde{a}_{m}-x_{m})|x,\overline{x}>=0, &  &  \\
&  & m>0 \\
(\widetilde{a}_{m}^{\dagger }+a_{m}-\overline{x}_{m})|x,\overline{x}>=0, &
&
\end{array}
\]
and the completeness condition
\[
\int \mathcal{D}x\,\mathcal{D}\overline{x}\,|x,\overline{x}><x,\overline{x}%
|=1
\]
The solution is
\begin{equation}
|x,\overline{x}>=\exp \left[ -\left( a^{\dagger },\widetilde{a}^{\dagger
}\right) -\frac{1}{2}\left( \overline{x},x\right) +\left( a^{\dagger
},x\right) +\left( \overline{x},\widetilde{a}^{\dagger }\right) \right] |0>,
\label{bosban}
\end{equation}
where the scalar product is defined as
$(a,b)=\sum_{m>0}a_{m}b_{m}$.
The general bosonic boundary state
can be written
\[
|B_{X}>=\int \mathcal{D}x\,\mathcal{D}\overline{x}\,e^{-S(x,\overline{x})}|x,%
\overline{x}>,
\]
where $S(x,\overline{x})$ is a bosonic boundary action, describing
the interaction of the open strings with background fields. When
the latter is zero one recovers the usual Neumann boundary state
\[
|N_{X}>=\int \mathcal{D}x\,\mathcal{D}\overline{x}\,|x,\overline{x}>=\exp
\left( a^{\dagger },\widetilde{a}^{\dagger }\right) |0>.
\]

According to Callan and Thorlacius \cite{Callan:1989mm}, terms depending on $%
x$ and $\overline{x}$ in the exponential in (\ref{bosban}) can
also be interpreted as the action of a quantum mechanical variable
defined at the string end point and interacting with external
sources given by the right and left moving creation operators.
(Notice that these last all commute with each other and can be
considered as c-number sources. They were actually used as
external sources in order to derive reparametrization invariance
Ward identities \cite{Freed:1993wz}, searching for conformal
points, i.e. phase transitions, of dissipative models.) The term,
quadratic in $x$ and $\overline{x}$, is just the CL term on the
circle
\begin{equation}
\frac{1}{2}\left( \overline{x},x\right) =-\frac{i}{4\pi }\int_{0}^{2\pi
}X_{+}(\sigma )\frac{dX_{-}(\sigma )}{d\sigma }d\sigma =-\frac{1}{32\pi
^{2}\alpha ^{\prime }}\int_{0}^{2\pi }d\sigma \int_{0}^{2\pi }d\sigma
^{\prime }\frac{X(\sigma )X(\sigma ^{\prime })}{\sin ^{2}\left( \dfrac{%
\sigma -\sigma ^{\prime }}{2}\right) },  \label{bosecircle}
\end{equation}
where, in the second identity $X_{\pm }(\sigma )$ are the positive
and negative frequency parts of $X(\sigma ).$ In the last identity
we have restored $\alpha ^{\prime }$ . This term is the bosonic
string action evaluated on the solution of the equation of motion
which is regular in the interior of the disk.

It is straightforward to show that, in the case of the fermionic string, the
same method leads to a non local fermionic action. To avoid complications
due to the zero modes which are present in the Ramond sector, we limit our
discussion to Neveu-Schwartz fermions, which are antiperiodic in the
variable $\sigma $. The fermionic boundary state is defined through the
equations
\begin{equation}
\begin{array}{lll}
(\psi _{m}^{\dagger }\pm i\widetilde{\psi }_{m}-\overline{\chi }_{m})|\chi ,%
\overline{\chi };\pm>=0, &  &  \\
&  & m>0 \\
(\widetilde{\psi }_{m}^{\dagger }\pm i\psi _{m}-i\chi _{m})|\chi ,\overline{%
\chi };\pm>=0, &  &
\end{array}
\label{ferbancond}
\end{equation}
and with the correct normalization  given by
\begin{equation}
|\chi ,\overline{\chi },\pm >=\exp \left[ \pm i\left( \psi ^{\dagger },%
\widetilde{\psi }^{\dagger }\right) -i\left( \overline{\chi },\chi \right)
+i\left( \psi ^{\dagger },\chi \right) \pm \left( \overline{\chi },%
\widetilde{\psi }^{\dagger }\right) \right] |0>.  \label{ferban}
\end{equation}
Following \cite{deAlwis:2001hi}, we have redefined in
(\ref{ferbancond}) and (\ref{ferban}) the variable $\chi $ and
$\overline{\chi }$, with respect to \cite{Abouelsaood:gd}, in
order to write the amplitude as a classical action. The general
fermionic boundary state reads
\[
|B_{\chi },\pm >=\int \mathcal{D}\chi \,\mathcal{D}\overline{\chi }%
\,e^{-S(\chi ,\overline{\chi })}|\chi ,\overline{\chi },\pm >,
\]
where, as in the bosonic case, $S(\chi ,\overline{\chi })$ is a
one-dimensional boundary action.

Extending the interpretation of \cite{Callan:1989mm} to the fermionic case,
terms depending on $\chi $ and $\overline{\chi }$ in the exponential in (\ref
{ferban}) can also be considered as the action of a one-dimensional
anticommuting variable defined at the string end point and interacting with
external sources. The fermionic non local term on the circle is given by
\begin{equation}
-i\left( \overline{\chi },\chi \right) =\frac{1}{8\pi ^{2}\alpha ^{\prime }}%
\text{P}\int_{0}^{2\pi }d\sigma \int_{0}^{2\pi }d\sigma ^{\prime }\frac{\chi
(\sigma )\chi (\sigma ^{\prime })}{\sin \left( \dfrac{\sigma -\sigma
^{\prime }}{2}\right) }.  \label{fermicircle}
\end{equation}
To compare (\ref{bosecircle}) and (\ref{fermicircle}) with the results of
Section 3, one has to make a string world-sheet duality transformation $%
\sigma \rightarrow \left( 2\pi /T\right) t$, remembering that world-sheet
fermions contribute an additional factor $\left( 2\pi /T\right) $ because
they have conformal weight $1/2.$ One then recovers eqs. (\ref{fulldissbos})
and (\ref{fulldissa}).

\section{$SL(2,R)$ symmetry and supersymmetry}
It was noticed in \cite{Callan:1989mm} that the non-local
Caldeira-Leggett
term on the real line is invariant only under the $SL(2,R)$ (or under $%
SU(1,1)$ for the theory on the circle) subgroup of the full
infinite group of reparametrizations. Full invariance would be
recovered only at critical points, if they exist. One can  see
that the fermionic action (\ref {fermidissi}) is invariant under
the same $SL(2,R)$ group. The key point is that $\chi $ has a
two-dimensional origin and as such has a conformal
dimension $1/2$. Then under a reparametrization $\tau \rightarrow \widetilde{%
\tau }=f(\tau )$ it transforms as $\widetilde{\chi }(\tau )=\left( f^{\prime
}(\tau )\right) ^{1/2}\widetilde{\chi }(\widetilde{\tau })$, then
\[
\int_{-\infty }^{\infty }d\tau \int_{-\infty }^{\infty }d\tau ^{\prime }%
\dfrac{\chi (\tau )\chi (\tau ^{\prime })}{\tau -\tau ^{\prime }}%
=\int_{-\infty }^{\infty }d\widetilde{\tau }\int_{-\infty }^{\infty }d%
\widetilde{\tau }^{\prime }\dfrac{\chi (\widetilde{\tau })\chi (\widetilde{%
\tau }^{\prime })}{\widetilde{\tau }-\widetilde{\tau }^{\prime
}}\left( \frac{f(\tau )-f(\tau ^{\prime })}{\left( \tau -\tau
^{\prime }\right) \left( f^{\prime }(\tau )\right) ^{1/2}\left(
f^{\prime }(\tau ^{\prime })\right) ^{1/2}}\right),
\]
which gives as condition for the invariance
\[
f(\tau )-f(\tau ^{\prime })=\left( \tau -\tau ^{\prime }\right) \left(
f^{\prime }(\tau )\right) ^{1/2}\left( f^{\prime }(\tau ^{\prime })\right)
^{1/2}.
\]
Taking the square of the two terms one recovers the relation which
guarantees the invariance of the CL action, and which establishes that $%
f(\tau )$ is a transformation of \ $SL(2,R).$

When we consider the full superstring action containing both the
bosonic and the fermionic terms there is a residual supersymmetry
which survives the integration over the oscillator modes. The
dissipative fermionic term (\ref {fermidissi}) adds to the CL term
(\ref{bosodissi}) coming from the integration on the bosonic
modes, producing a non-local supersymmetric dissipative term.
After a trivial rescaling of the fields, it can be written as
\begin{equation}
S_{diss}=\int dt_{1}\int dt_{2}\left( \frac{q(t_{1})q(t_{2})}{%
(t_{1}-t_{2})^{2}}-\dfrac{\chi (t_{1})\chi (t_{2})}{t_{1}-t_{2}}\right) ,
\label{superdisscomp}
\end{equation}
which is invariant under the supersymmetry transformation
\begin{equation}
\begin{array}{lll}
\delta q=\epsilon \chi , &  & \delta \chi =\epsilon \dfrac{dq}{dt},
\end{array}
\label{susy}
\end{equation}
where $\epsilon $ is a real constant anticommuting parameter. In fact, (\ref
{superdisscomp}) can be easily written using superspace formalism. Let us
introduce a Grassmann \ coordinate $\theta $ and the supervariable
\[
X(t,\theta )=q(t)+\theta \chi (t).
\]
The superspace possesses the symmetry
\[
\begin{array}{lllllll}
\delta _{\varepsilon }t=\varepsilon t, &  & \delta _{\varepsilon }\theta
=\varepsilon , &  & \varepsilon ^{2}=0, &  & \{\varepsilon ,\theta \}=0.
\end{array}
\]
The supersymmetry generator is
\[
Q=\partial _{\theta }-\theta \partial _{t}.
\]
Then the action (\ref{superdisscomp}) can be obtained integrating over the $%
\theta ^{\prime }$s the following expression
\begin{equation}
S_{diss}=\int dt_{2}dt_{1}d\theta _{2}d\theta _{1}\dfrac{X(t_{1},\theta
_{1})X(t_{2,}\theta _{2})}{z_{12}}  \label{superdiss}
\end{equation}
where $z_{12}=t_{1}-t_{2}+\theta _{1}\theta _{2}.$  Let us notice
that there is an obvious generalization of the non-local
supersymmetric action (\ref{superdiss}) \textit{i.e.}
\begin{eqnarray}
S_{{}} &=&\int dt_{2}dt_{1}d\theta _{2}d\theta _{1}X(t_{1},\theta
_{1})W(z_{12})X(t_{2,}\theta _{2})=  \label{supergen} \\
&&\int dt_{1}\int dt_{2}\left( q(t_{1})W^{\prime }(t_{1}-t_{2})q(t_{2})-\chi
(t_{1})W(t_{1}-t_{2})\chi (t_{2})\right) ,  \nonumber
\end{eqnarray}
with $W(t_{1}-t_{2})=-W(t_{2}-t_{1}).$ In fact, another example of
action of the form (\ref{supergen}) is provided by the theory on
the circle obtained by adding the action (\ref{fulldissp}) of
fermions periodic in time to the CL term (\ref{fulldissbos}).

Actually, one can check that the action (\ref{superdiss}) is
invariant under the larger group of super-conformal
transformations
\begin{eqnarray}
t &\rightarrow &\dfrac{at+b}{ct+d}+\frac{-\delta t+\varepsilon }{(ct+d)^{2}}%
\theta  \label{realmoebius} \\
\theta &\rightarrow &\frac{\theta }{ct+d}+\frac{-\delta t+\varepsilon }{ct+d}
\nonumber
\end{eqnarray}
where $a,b,c,d$ are real constants and $\delta ,\varepsilon $ are
real anticommuting constants satisfying the relation
$ad-bc=1-\delta \varepsilon $. These transformations are the real
subgroup of the super-Moebius
transformation acting on a complex commuting variable $z$ and a couple $%
\theta ,\overline{\theta }$ of Grassmann variables \cite{Ezawa:1996fh}, (see
also \cite{Verlinde:2004gt}). The corresponding Lie algebra contains three
bosonic charges: the Hamiltonian $H$, which is the generator of time
translations, the time dilatation generator $D$, and $K$ which generates
conformal transformations. The remaining anticommuting charges are the
generator of supersymmetry, $Q$, and the generator of special (conformal)
transformations, $S$. Explicitly, following the notations of ref. \cite
{Clark:2001zv}, the algebra is characterized by the following non vanishing
(anti-)commutation relations
\begin{equation}
\begin{array}{ccccc}
\lbrack H,D]=iH &  & [H,K]=2iD &  & [D,K]=iK \\
&  &  &  &  \\
\{Q,Q\}=H &  & [D,Q]=-\frac{i}{2}Q &  & [K,Q]=\frac{i}{2}S \\
&  &  &  &  \\
\{S,S\}=K &  & [D,S]=iS &  &  \\
&  &  &  &  \\
\{Q,S\}=-D &  & [H,S]=-iQ &  &
\end{array}
\label{realalgebra}
\end{equation}
The same algebra (\ref{realalgebra}), was considered in the study
of the geometry of superconformal quantum mechanics made in ref.
\cite{Michelson:1999zf}, where it was classified as the
$\mathcal{N}=1B$ extension of the Poincar\`{e} subalgebra to
$Osp(2|1)$ superconformal algebra. The (super-)conformal
invariance of the dissipative term makes it play a dominant role in
determining the long range physical properties of a dissipative
system.

\section{Supersymmetric dissipative quantum mechanics and other boundary
interactions}
The non-local supersymmetric dissipative term can be
added to the usual supersymmetric kinetic term, which is also
invariant under (\ref{susy})
\begin{equation}
S_{kin}=-\frac{1}{2}\int dt\int d\theta \left( D^{2}XDX\right) =\frac{1}{2}%
\int dt\left( \left( \frac{dq}{dt}\right) ^{2}-\chi \frac{d\chi }{dt}\right)
\label{skin}
\end{equation}
where $D=\partial _{\theta }+\theta \partial _{t}$, and
$\{Q,D\}=0$. The system described by the total action
$S_{kin}+S_{diss}\ $is thus a supersymmetric extension of
dissipative quantum mechanics. Notice however that the kinetic
term breaks the $SL(2,R)$ invariance of (\ref{superdiss}), and
(super-)conformal invariance is lost. The physical interpretation
of the model is better understood if we extend it to the case of
more then one (super-)variable. Let us consider for instance the
case of $D=2k$ supervariables $X^{\mu }(t,\theta )=q^{\mu
}(t)+\theta \chi ^{\mu }(t)$, with $\mu =1,...,D.$ It is well
known that the path integral on only the commuting variable
$q^{\mu }$ gives the propagator of a scalar field in $D$
dimension, while path integral on only the anticommuting variable
$\chi ^{\mu }$ gives a pseudoclassical Lagrangian representation
of the spin (for simple discussions see for instance
\cite{Polyakov}, \cite{Peskin:1994be}). This last fact is better
understood when (\ref{skin}) is considered as the starting point
of canonical quantization. As the kinetic term for the $\chi ^{\mu
}$ in (\ref{skin}) is linear in time derivatives, half of the
anticommuting variables are the conjugate momenta of the remaining
half. Quantization leads then a to the irreducible $2^{D/2}$
spinor representation of the $D$-dimensional Grassmann algebra.
(This is why we considered $D$
even. A discussion about the partition function for a single anticommuting $%
\chi $ can be found in \cite{Witten:1998cd}). When computing the amplitude
from $(q^{\mu },\chi ^{\mu })$ to $(q^{\prime \mu },\chi ^{\mu })$ (the
quantity $\chi ^{\mu }$ remains constant due to the equation of motion $%
d\chi ^{\mu }/dt=0$), one obtains the propagator for \ a Dirac particle \cite
{Brink:uf}.

One can add another local supersymmetric term to introduce a potential into
the action. Its form is familiar from recent studies on tachyon condensation
in superstring theory \cite{Witten:1998cd}, \cite{Harvey:2000na}, \cite
{Kutasov:2000aq}.
\begin{equation}
S_{T}=\int dt\int d\theta \left( \Gamma D\Gamma +\Gamma T(X)\right)
\label{stach}
\end{equation}
where the scalar (super-)potential $T(X)$ represents in string
theory the tachyon field and $\Gamma (t,\theta )=\eta (t)+\theta
F(t)$ is a new anticommuting supervariable living on the string
end-point (whose presence in string theory is needed in order to
have correct space-time properties of the tachyon vertex
operator). Again, one has to restrict to (NS)
antiperiodic $\eta$, to avoid problems with zero-modes. Integration on $%
\theta $, and functional integration on the auxiliary field $F,$ lead to the
following expression
\[
S_{T}=-\frac{\alpha ^{\prime }}{8}\int dt\int dt^{\prime }\left[ \chi (t)%
\frac{\partial }{\partial q}T(q(t))\right] \,sign(t-t^{\prime
})\,\left[ \chi
(t^{\prime })\frac{\partial }{\partial q}T(q(t^{\prime }))\right] -%
\frac{1}{4}\int dt\,T(q(t))^{2}
\]
where we have restored $\alpha ^{\prime }$. Notice that the scalar
potential for $q$ is positive definite.

Let us make some general remarks on the one-dimensional models
obtained from an action containing (\ref{skin}), (\ref{superdiss})
and (\ref {stach}). Conformal invariance is lost in presence of
(\ref{skin}) and (\ref {stach}) and one should then study the
properties of the model under the renormalization group flow. In
particular, one can hope that full conformal invariance is
restored at some non trivial fixed point. The behavior of the
various terms under scaling transformations is different. The
non-local dissipative term is scaling invariant, while the kinetic
term is an irrelevant operator and the term containing the tachyon
potential is marginal or relevant. All these facts were well known
in the case of the purely bosonic case \cite{Schmid:1983ff},
\cite{Fisher:1983ff}. For instance, for what concerns the long
range (IR) behavior of the model, the kinetic terms (which is an
irrelevant operator) plays just the role of an UV cut-off and can
be discarded in favor of simpler cut-off procedures in the course
of renormalization analysis. It is then the dissipative term,
which is scale-invariant, to play, in the quantum mechanical
model, the role of inverse propagator, and so is the
one-dimensional analogous of the laplacian operator in two
dimensions. When the model contains more than one variable, one can
add other interesting terms to the action. For instance, in the
case of $D=2$, one can couple the system to an external magnetic
field obtaining models similar to the dissipative Hofstadter
model, which has been widely studied for its remarkable duality
properties \cite{Callan:1991da}, \cite {Callan:1992vy}. It would
be interesting to study possible dualities of the supersymmetric
version of that model.

\section{Conclusions and outlook}

With hindsight the derivation of the CL model from the dynamics of
the open string end-point may seem not surprising, since, with its
infinite number of oscillation modes, the open string just
provides a model of the bath of oscillators required in the
Caldeira-Leggett approach to dissipation. This point of view is
advocated in \cite{Affleck:2000ws}, where it is stated that when
the oscillator spectral weight vanishes linearly at low
frequencies, \textit{i.e.} in the case of Ohmic dissipation, the
set of oscillators may be represented by a (1+1) dimensional
quantum conformal field theory of free massless bosons living on a
fictitious half-line. An alternative approach to dissipation can
be found in \cite{Fujikawa:1998bh}, \cite{Fujikawa:1998hq} and
\cite{Terashima:1999xp}. In those papers a reformulation of the
fluctuation-dissipation theorem was introduced in such a manner
that the basic idea of simulating the dissipative environment by
an infinite number of bosonic oscillator was realized without
manifestly introducing them. It is interesting that this
formulation was also extended to fermions. In particular in
ref.\cite{Terashima:1999xp}, these ideas where applied to a
fermionic detector (a single quantum mechanical fermionic degree
of freedom) coupled to a Dirac field which plays the role of a
fermionic dissipative environment.

Let us make a remark on a possible extension from quantum
mechanics to field theory. The most obvious generalization of the
non-local fermionic term (\ref{fermidissi}) to a Dirac spinor
field in Minkowski four-dimensional spacetime, would lead to the
action
\[
\mathbf{S}=\displaystyle{\int d^{4}x\,\bar{\psi}(x)(i\kern-4pt
\not{\hbox{ \kern-6pt $\partial$}-m}) \psi (x)+{\frac{i\eta }{\pi
}}\int d^{3}x\int dtdt^{\prime
}\,\bar{\psi}(t,\mathbf{x})\,{\frac{1}{t-t^{\prime }}}\,\psi
(t^{\prime },\mathbf{x}).}
\]
It reduces to quantum mechanics if we consider fields independent
of the spatial coordinates (which could be of interest in
cosmological applications). A similar action has been proposed 
as CPT violating but Lorentz invariant model of fermions
(neutrinos)\cite{Barenboim:2002tz}: CPT theorem would be violated
due to the non locality of the action. However, propagators are
not causal and also not covariant 
\cite{Barenboim:2002tz,Greenberg:2003ks}. It may
be that this model could be given
a new interpretation in light of our results: our understanding is
that i) the non-local term could be considered not as a mass term
but rather as the effect of an interaction with a bath of
fermionic oscillators, and that ii) as a dissipative term, it
provides a rationale for introducing irreversibility into the
system. Thus we think that our interpretation is consistent with
the criticism of this model expressed in \cite{Greenberg:2003ks}.

Let us finally stress the main result of this paper. We have
constructed new dynamical models extending the fermionic particle
model described by the lagrangian (\ref{skin}): we have shown how
to include dissipative effects by means of a straightforward
extension of the CL term. Once again, string theory has proven to
be a valuable tool to derive these results, following the seminal
ideas of Callan and Thorlacius.

\section*{Acknowledgments}

We thank G. Maiella for fruitful discussions about CL and CT papers. L.C. is
partially supported by the EC RTN Programme HPRN-CT-2000-00131 and by MIUR
project 2003-023852. The work of G.D. is supported in part by TMR,
EC-Contract No. ERBFMRX-CT980169 (EURODA$\Phi $NE).


\begin{thebibliography}{99}
\bibitem{Caldeira:1982uj}  A.~O.~Caldeira and A.~J.~Leggett,
Annals Phys.\ \textbf{149}, 374 (1983).


\bibitem{Callan:1989mm}  C.~G.~Callan and L.~Thorlacius,
Nucl.\ Phys.\ B \textbf{329}, 117 (1990).


\bibitem{Fradkin:1985qd}  E.~S.~Fradkin and A.~A.~Tseytlin,
Phys.\ Lett.\ B \textbf{163}, 123 (1985).


\bibitem{Abouelsaood:gd}  A.~Abouelsaood, C.~G.~.~Callan, C.~R.~Nappi and
S.~A.~Yost, 
Nucl.\ Phys.\ B \textbf{280}, 599 (1987).


\bibitem{Polchinski:1987tu}  J.~Polchinski and Y.~Cai,
Nucl.\ Phys.\ B \textbf{296}, 91 (1988).


\bibitem{Callan:1988wz}  C.~G.~.~Callan, C.~Lovelace, C.~R.~Nappi and
S.~A.~Yost, 
Nucl.\ Phys.\ B \textbf{308}, 221 (1988).


\bibitem{Iengo:1997hz}  R.~Iengo and G.~Jug,
arXiv:hep-th/9702021.

\bibitem{Barenboim:2002tz}  G.~Barenboim and J.~Lykken,
Phys.\ Lett.\ B \textbf{554}, 73 (2003) [arXiv:hep-ph/0210411].


\bibitem{Greenberg:2003ks}  O.~W.~Greenberg, 
Phys.\ Lett.\ B \textbf{567}, 179 (2003) [arXiv:hep-ph/0305276].

\bibitem{weiss}  U. Weiss, \textit{Quantum Dissipative Systems}, World
Scientific, 1993.


\bibitem{Green:sp}  M.~B.~Green, J.~H.~Schwarz and E.~Witten, \textit{%
Superstring Theory. Vol. 1: Introduction}, Cambridge University Press, 1987.


\bibitem{Freed:1993wz}  D.~E.~Freed,
Nucl.\ Phys.\ B \textbf{424}, 628 (1994) [arXiv:hep-th/9309116].


\bibitem{deAlwis:2001hi}  S P. de Alwis,
Phys. Lett. B \textbf{505}, 215 (2001) [arXiv:hep-th/0101200].

\bibitem{Ezawa:1996fh}  K.~Ezawa and A.~Ishikawa,
Phys.\ Rev.\ D \textbf{56}, 2362 (1997) [arXiv:hep-th/9612031].


\bibitem{Verlinde:2004gt}  H.~Verlinde,
arXiv:hep-th/0403024.


\bibitem{Clark:2001zv}  T.~E.~Clark, S.~T.~Love and S.~R.~Nowling,
Nucl.\ Phys.\ B \textbf{632}, 3 (2002) [arXiv:hep-th/0108243].


\bibitem{Michelson:1999zf}  J.~Michelson and A.~Strominger,
Commun.\ Math.\ Phys.\ \textbf{213}, 1 (2000) [arXiv:hep-th/9907191].


\bibitem{Brink:uf}  L.~Brink, P.~Di Vecchia and P.~S.~Howe,
Nucl.\ Phys.\ B \textbf{118}, 76 (1977).

\bibitem{Polyakov}  A. M. Polyakov, \textit{Gauge Fields and Strings, }%
Harwood Academic Publishers, 1987.


\bibitem{Peskin:1994be}  M.~E.~Peskin, \textit{Spin, mass, and symmetry}
Lectures given at 21st Annual SLAC Summer Institute on Particle Physics:
Spin Structure in High Energy Processes, Stanford, CA, 26 Jul - 6 Aug 1993.
[arXiv:hep-ph/9405255].


\bibitem{Witten:1998cd}  E.~Witten, 
JHEP \textbf{9812}, 019 (1998) [arXiv:hep-th/9810188].


\bibitem{Harvey:2000na}  J.~A.~Harvey, D.~Kutasov and E.~J.~Martinec,
[arXiv:hep-th/0003101].


\bibitem{Kutasov:2000aq}  D.~Kutasov, M.~Marino and G.~W.~Moore,
[arXiv:hep-th/0010108].


\bibitem{Schmid:1983ff}  A.~Schmid,
Phys.\ Rev.\ Lett. \textbf{51}, 1506 (1983).


\bibitem{Fisher:1983ff}  M.P.A. Fisher and W.~Zwerger,
Phys.\ Rev.\ B \textbf{32}, 6190 (1985).


\bibitem{Callan:1991da}  C.~G.~.~Callan and D.~Freed,
Nucl.\ Phys.\ B \textbf{374}, 543 (1992) [arXiv:hep-th/9110046].


\bibitem{Callan:1992vy}  C.~G.~Callan, A.~G.~Felce and D.~E.~Freed,
Nucl.\ Phys.\ B \textbf{392}, 551 (1993) [arXiv:hep-th/9202085].

\bibitem{Affleck:2000ws}  I.~Affleck, M.~Oshikawa and H.~Saleur,
Nucl.\ Phys.\ B \textbf{594}, 535 (2001) [arXiv:cond-mat/0009084].


\bibitem{Fujikawa:1998bh}  K.~Fujikawa,
Phys.\ Rev.\ E \textbf{57}, 5023 (1998) [arXiv:hep-th/9802025].


\bibitem{Fujikawa:1998hq}  K.~Fujikawa and H.~Terashima,
Phys.\ Rev.\ E \textbf{58}, 7063 (1998) [arXiv:hep-th/9809048].


\bibitem{Terashima:1999xp}  H.~Terashima,
Phys.\ Rev.\ D \textbf{60}, 084001 (1999) [arXiv:hep-th/9903062].


\end{thebibliography}
\end{document}